\documentclass[12pt, a4paper]{article}

\usepackage[latin1]{inputenc}
\usepackage{vmargin}
\usepackage{amsfonts}
\usepackage{amssymb}
\usepackage{graphicx}
\usepackage{amsmath}
\usepackage{amsthm}
\usepackage{setspace}
\usepackage{multirow}
\usepackage{hyperref}
\usepackage{cases}
\usepackage[round]{natbib}

\usepackage{stackengine}
\def\yenrule{\rule{1.3ex}{.1ex}}
\def\textyen{\renewcommand\stacktype{L}\stackon[.4ex]{\stackon[.65ex]{Y}{\yenrule}}{\yenrule}}

\newtheorem{theorem}{Theorem}[section]

\newtheorem{assumption}{Assumption}[section]

\let\Item\item

\renewcommand\enddescription{\endlist\global\let\item\Item}

\def\BAL{\mbox{Bal}}
\def\LT{\mbox{LTick}}
\def\ambig{{\tiny Ambiguous}}

\usepackage[bottom]{footmisc}
\setmarginsrb{2cm}{3cm}{2cm}{2cm}{0cm}{0cm}{0cm}{0.5cm}
\title{How to predict the consequences of a tick value change? Evidence from the Tokyo Stock Exchange pilot program}

\author{Weibing Huang$^{1}$, Charles-Albert Lehalle$^{2}$  and Mathieu Rosenbaum$^{1}$\\$~~$\\
$^{1}${\small LPMA, University Pierre and Marie Curie (Paris 6)}\\
$^{2}${\small Capital Fund Management, Paris and CFM-Imperial College Institute, London}}

\date{\today}

\begin{document}

\maketitle

\begin{abstract}
\noindent The tick value is a crucial component of market design and is often considered the most suitable tool to mitigate the effects of high frequency trading.
The goal of this paper is to demonstrate that the approach introduced in \citet*{dayri2012large} allows for an ex ante assessment of the consequences of a tick value change on the microstructure of an asset. To that purpose, we analyze the pilot program on tick value modifications started in 2014 by the Tokyo Stock Exchange in light of this methodology. We focus on forecasting the future cost of market and limit orders after a tick value change and show that our predictions are very accurate. Furthermore, for each asset involved in the pilot program, we are able to define (ex ante) an optimal tick value. This enables us to classify the stocks according to the relevance of their tick value, before and after its modification. 

% In this work, we study the effects of tick value reduction for large tick assets, for which the bid-ask spread is almost always equal to one tick. For these assets, we introduce the notion of implicit spread, using the high frequency indicator $\eta$ from the model with uncertainty zones. The equivalency between the implicit spread and the conventional bid-ask spread for large tick asset is shown by both arguments of market efficiency and empirical studies. Moreover, these analysis enable us to predict the effect of tick value changes for large tick assets. The prediction formulas are tested using the data of the Japanese tick size reduction pilot program, and are shown to be very consistent in term of their prediction quality. 
\end{abstract}

% a simple reasoning, supported by strong empirical evidences allows to build an estimate of the relative cost of market orders to limit orders. This indicator of the quality of market microstructure is deeply linked to the tick size [ref]. It allows us to classify Japanese stocks into three categories: costly for market makers, costly for investors, and others.
% during the Japanese tick size experiment, N1 stocks changed of category during the first phase, and N2 during the second one. The reduction of the tick size did not systematically improved the cost of market orders, usually associated investors trading costs. Nevertheless we identified some cases for which the tick size reduction improved the cost of market orders, and hence most probably decreased the cost for institutional investors. 
% We show how this indicator is far more subtle than any other measurement, like the ones based on the bid-ask spread in tick size of in basis points, or on the market depth. Moreover, based on our model of the cost of market vs limit orders, we could predict with a very good accuracy (in basis points) the new value of the bid-ask spread after the tick changes of the Japanese experiment.
% The accuracy of our predictions underlines the important role of the transaction rates (and more specifically the ratio of the rate of continuing trades vs the one of alternating ones) in the determination of the optimal tick size.

\section{Introduction}

On January 14, 2014, the Tokyo Stock Exchange (TSE) launched the first phase of its pilot program on tick value\footnote{The tick value is the minimum price variation allowed for an asset on a given market.} modifications, reducing the tick value of the TOPIX 100 index stocks priced above \textyen3000 by approximately 90\% (see Section \ref{secdata} for more details on this pilot program). The second phase was implemented on July 22, 2014, targeting a sub-Yen tick value reduction for stocks of the TOPIX 100 index priced below \textyen5000. The third phase of this program is expected to start in September 2015, when a new tick value table should be announced after the evaluation of the effects of the tick value reductions in the first two phases. \\

The tick value is probably the most relevant tool that can be used by exchanges and regulators in order to improve the trading quality and the robustness of the market structure, see \cite{citeulike:12047995}. Compared with other more controversial proposals, such as imposing a minimum resting time for orders to remain valid or using frequent batch auctions, setting a suitable tick value is in general considered to be a better way to control the growing activity of high frequency traders, which accounts nowadays for more than 40\% of the total volume on equity markets. Indeed, a tick value change induces very little cost and is easily reversible if the outcome does not meet the market designer's expectations.\\

The Tokyo Stock Exchange is not alone in its search for better tick values. In the United States, on August 26, 2014, the Securities and Exchange Commission announced a program aiming at widening tick values for stocks with small capitalization. A targeted 12 months pilot experiment will be implemented to assess the effects of such changes. In Europe, in May 2014, the European Securities and Markets Authority released a MiFID2/R discussion paper which proposes two options for a new harmonized tick value regime to be introduced across all trading venues. These two options are currently debated by European regulators.  
% Both options takes into account the liquidity and a target bid-ask spread. However, they differ in the scope of the tick value regime table (QUE VEUX TU DIRE PAR LA), the quantification of the liquidity levels and the target value for the bid-ask spread.
\\

Before the pilot program, most Japanese stocks were typical examples of {\it large tick assets}, that is assets whose bid-ask spread is almost always equal to one tick. The tick value being the lower bound for the bid-ask spread, when it is too large, the cost of market orders becomes very significant. This not only damages liquidity takers but also the ``slow'' liquidity providers who suffer from the intensification of the speed competition for gaining time priority in the order book queues, see \citet*{moallemi}. Although it is quite commonly accepted that it is preferable to reduce the tick value for these large tick assets, finding the appropriate tick value remains a very intricate problem. Indeed, most of the numerous works about the consequences of a tick value change are empirical and focus on the outcomes of this market design modification in an {\it ex post} basis, see for example \citet*{lau1995reducing}, \citet*{ahn1996tick}, \citet*{bacidore1997impact}, \citet*{bessembinder2000tick}, \citet*{goldstein2000eighths}, \citet*{chung2001order}, \citet*{chung2002tick}, \citet*{bourghelle2004markets} and \citet*{wu2011impacts}.\\

These studies have clearly shown that a change in the tick value may lead to significant implications for the bid-ask spread, the available volume in the order book and many other microstructural quantities. However, very few quantitative tools exist for predicting {\it ex ante} these effects. Consequently, exchanges and market regulators often rely on the trial and error approach in order to set appropriate tick values. The Japanese pilot experiment, in which the tick value reduction program is conducted in three phases, is one of such examples. Indeed, the effects of tick value changes in the first two phases are evaluated ex post to help the design of a new tick value table to be implemented in the last phase.\\

In \citet*{dayri2012large}, the authors build a quantitative approach towards solving the crucial problems of forecasting the consequences of a tick value change and determining an optimal tick value. To that purpose, based on the model with uncertainty zones introduced in \citet*{robert2011new}, they use the key microstructural indicator $\eta$ (half the ratio between price continuations and alternations, see Section \ref{sec:model}) which summarizes the high frequency features of an asset. The paramount importance of the parameter $\eta$ is due to the fact that there is a one to one bijection between its value and the cost of market and limit orders. We recall this connection in details in Section \ref{sec:model}. Hence, measuring $\eta$ allows us to classify stocks according to whether they are profitable for market makers or rather balanced. Furthermore, being able to predict the consequences of a tick value change on $\eta$ means one can anticipate the new microstructural costs induced by this tick value modification, which is precisely what exchanges and regulation authorities need. Such predictions are possible using the approach in \citet*{dayri2012large} where explicit forecasting formulas for the parameter $\eta$ are provided. Moreover, the way to set a tick value leading to an optimal $\eta$ is also established (see Section \ref{sec:model} for our definition of optimality).\\

In this work, our goal is to show that the theoretical forecasting formulas in \citet*{dayri2012large} do enable us to predict ex ante the consequences of a tick value change on the microstructure of an asset, notably on the trading costs. To demonstrate this, we use 18 months of tick by tick market data from the TSE, including the whole year 2014 when the pilot program is in place. Very accurate results are obtained for the prediction of the parameter $\eta$. Thus, the  approach in \citet*{dayri2012large} is proved to be indeed very helpful in both predicting the consequences of a tick value change and choosing an optimal tick value for large tick assets. \\

The paper is organized as follows. We recall in Section \ref{sec:model} the reading of the model with uncertainty zones as a mean to quantify the average cost of market and limit orders using the crucial microstructural indicator $\eta$. At the end of this section, we give the prediction formula for $\eta$ after a tick value modification. Hence we provide a way to predict the change in the cost of market and limit orders induced by such modification. In Section \ref{empstudy}, we consider the TSE pilot experiment on tick values. First, before the start of the program, we classify Japanese assets in two categories: stocks with costly market orders and stocks with balanced costs between market and limit orders. Note that the situation of costly limit orders is very unlikely. Indeed, in that case, market makers would increase their spread, what they can always do. Then we apply in the same section the forecasting methodology presented in Section \ref{sec:model}. In particular, we predict whether a stock will change category after the tick value modification. We conclude in Section \ref{conclu}.

%In Section 2, we introduce briefly the model with uncertainty zones and the quantitative tools for studying the tick value effects. In the end of Section 2, we present our method for predicting the effects of tick value reduction for large tick assets, and introduce the notion of optimal tick value, which minimises the transaction cost of market orders and the bid-ask spread in units of tick value at the same time. In Section 3, the prediction results are validated in our empirical studies using the data from the Japanese pilot program, and are shown to be very consistent. 

\section{Cost of trading and high frequency price dynamics}
\label{sec:model}
%\section{The High Frequency Indicator $\eta$}

\subsection{The model with uncertainty zones: When the tick prevents price discovery}

\begin{figure}
\centering
\includegraphics[width=\textwidth]{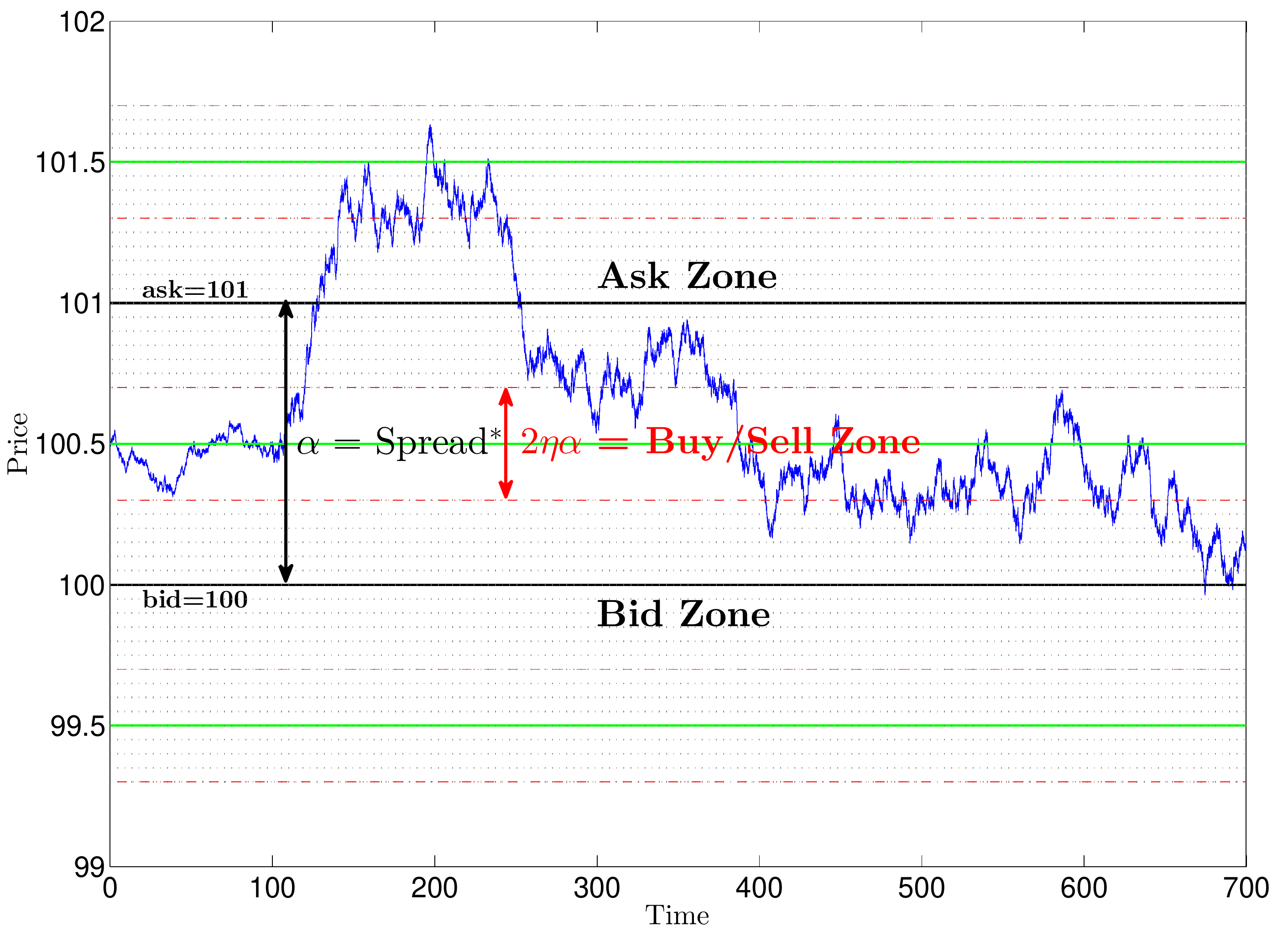}
\caption{The three different zones when the bid-ask is 100-101 and the tick value is equal to one. The red dotted lines are the limits of the uncertainty zone. The uncertainty zone inside the spread is the buy/sell zone. The upper dotted area is the ask zone and the lower dotted area is the bid zone.}\label{fig:Uncertainty_Shade3}
\end{figure}

The model with uncertainty zones, introduced in \citet*{robert2011new}, is a high frequency model for the transaction prices of a large tick asset. It reproduces most macroscopic and microscopic stylized facts of price dynamics and is very suitable for the analysis of the role of the tick value in determining the microstructural features of an asset. This model assumes the existence of a latent efficient price $X_t$, typically a martingale, and states that a transaction can occur at a given price level (on the tick grid) only provided this price level is close enough to the efficient price. This proximity is quantified by the parameter $\eta$: the distance between the potential transaction price and the efficient price has to be smaller than $\alpha/2+\eta\alpha$, with $\alpha$ the tick value of the asset. Thus, for a large tick asset, assuming the efficient price lies within the one tick bid-ask spread $[b, b+\alpha]$, we have $\eta\in[0,1/2]$ and obtain three zones for the efficient price: 
\begin{itemize}
\item If it lies between $b$ and $b+\alpha(1/2-\eta)$, transactions can only occur on the bid side (bid zone). 
\item If it lies between $b+\alpha(1/2+\eta)$ and $b+\alpha$, transactions can only occur on the ask side (ask zone). 
\item If it lies between $b+\alpha(1/2-\eta)$ and $b+\alpha(1/2+\eta)$, transactions can occur both on the bid and on the ask side (buy/sell or uncertainty zone).
\end{itemize}
These three zones are summarized in Figure \ref{fig:Uncertainty_Shade3}.

\paragraph{Estimation of $\eta$}

The parameter $\eta$ can be very easily estimated as follows. We define an alternation (resp. continuation) as a transaction price jump of one tick whose direction is opposite to (resp. the same as) the one of the preceding transaction price jump. Let $N^{(a)}$ and $N^{(c)}$ be respectively the number of alternations and continuations during the period $[0,t]$. The estimator of $\eta$ over $[0,t]$ is simply given by
$$
\widehat{\eta} = \frac{N^{(c)}}{2N^{(a)}}.$$
Theoretical properties for this estimator are established in \citet*{robert2012volatility}. Note that in Section \ref{empstudy}, the estimated values of $\eta$ over given time periods of several months will be given by the averages of the daily estimations of $\eta$ over all the days of the periods.\\

\subsection{Perceived tick size and cost of market orders}

\begin{figure}
\centering
\includegraphics[width=\textwidth,height=8cm]{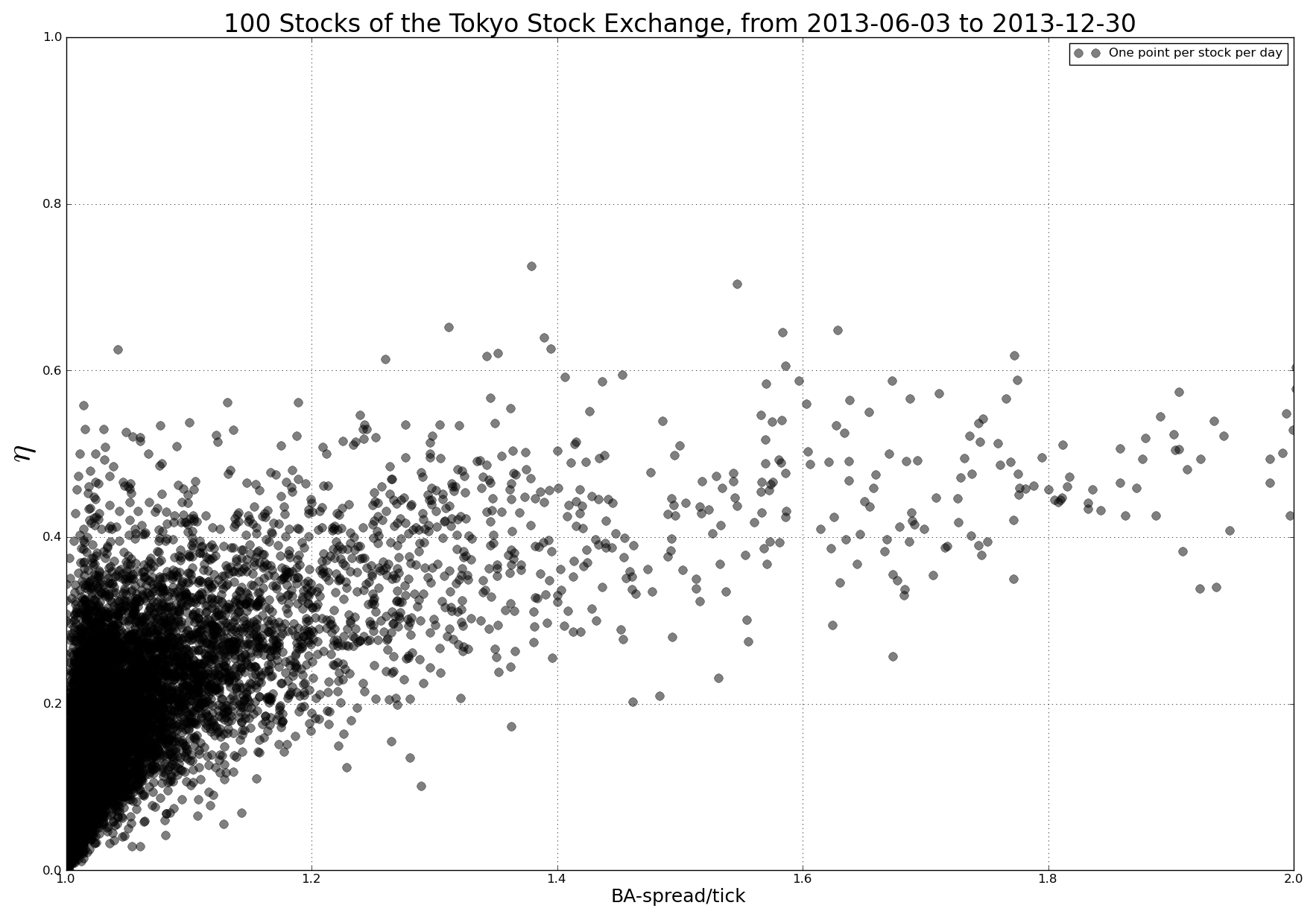}
\caption{Daily values of $\eta$ for all the assets of the TOPIX 100 index, for every day so that the daily average spread is smaller than two ticks, from 2013, June 3 to 2013, December 30.}\label{fig:etaspread}
\end{figure}

The parameter $\eta$ controls the width of the uncertainty zones (which is $2\eta\alpha$ ; when the efficient price is inside this zone, investors cannot clearly decide if it is more relevant to buy or sell) and measures the bouncing intensity of the transaction price due to the existence of the tick value. It can actually be seen as an indicator for the perceived tick size of a large tick asset: A very small $\eta$ ($\eta\ll 0.5$) means that for market participants, the tick value appears much too large (in such case, it is necessary to be sharp in term of estimation of the efficient price to know if it is reasonable to buy or sell at a given time), while a $\eta$ close to $1/2$ is synonym of a suitable tick value (in such situation, the uncertainty zones almost correspond to the tick grid). To understand this, consider a market order of unit volume at price $P_t$ at time $t$. Its cost with respect to the efficient price is $P_t - X_t$. For a large tick asset, the average cost of such market order can be computed and is equal to
\begin{equation*}
\alpha/2 - \eta \alpha,
\end{equation*}
see \citet*{dayri2012large}. The quantity $\alpha/2 - \eta \alpha$ is non negative provided $\eta\leq 0.5$, a condition which is almost systematically satisfied by estimated values of $\eta$ on large tick assets, see Figure \ref{fig:etaspread}. Indeed, it would otherwise mean that on average, market makers lose money. This is not really possible since in such situation, they would simply increase their spread, what they can always do. Thus, when $\eta < 0.5$, market takers have to pay a fixed positive cost in order to get liquidity, while market makers gain profit by placing limit orders\footnote{Of course this is true only considering the aggregated group of market makers. In practice, the gains of individual market makers are often small since there
are several of them and the queues at the best bid and ask levels are quite long.}. This cost paid by liquidity takers is given by $\alpha/2 - \eta \alpha$. Consequently, for large tick assets, the classical efficiency rule that assumes zero cost for both market and limit orders becomes irrelevant.\\

\subsection{Implicit bid-ask spread and cost of limit orders}

Within bid-ask quotes of the form $[b,b+\alpha]$, the width of the uncertainty zone represents the range of values for the efficient price $X_t$ where  transactions can occur both at the best bid and the best ask side. The size of this range is $2\eta\alpha$. Therefore, it is natural to view the quantity $2\eta\alpha$ as an {\it implicit spread}. This idea is fully supported by the regression analysis in \citet*{dayri2012large}, which shows empirically that for large tick assets, $\eta\alpha$ is proportional to the volatility per trade: 
\begin{equation}\label{volpertrade}
\eta\alpha\sim c\frac{\sigma}{\sqrt{M}},
\end{equation}
where $\sigma$ denotes the square root of the daily integrated variance of the price, $M$ the number of transactions per day and $c$ a constant around one. Yet it is well-known that for small tick assets, for which the (conventional) spread can evolve freely and is not artificially bounded from below by the tick value, the average spread is proportional to the volatility per trade, see \citet*{madhavan1997security} and \citet*{wyart2008relation}. This confirms that $2\eta\alpha$ can be interpreted as an implicit spread for a large tick asset.\\

Actually the fact that for small tick assets, the average spread $S$ is proportional to the volatility per trade simply comes from the efficiency condition stating that market makers make on average zero profit due to competition. More precisely, to derive the spread-volatility per trade relation, let us consider a dichotomy between market makers using limit orders and market takers using market orders. In that case, the average profit and loss per trade of a typical market making strategy, which can be understood as that of a limit order, is essentially equal to $S/2-c\sigma/\sqrt{M}$, see \citet*{wyart2008relation}. Therefore, the efficiency assumption implies 
$$\frac{S}{2}\sim c\frac{\sigma}{\sqrt{M}}.$$
In the case of a large tick asset, for which $S=\alpha$, as seen in the previous subsection, market orders are costly, their cost being on average $\alpha/2 - \eta\alpha$. Therefore the profit and loss of market makers, which is still $S/2-c\sigma/\sqrt{M}$, is no longer zero. Indeed, it is precisely the cost paid by market takers. Consequently, we get
$$S/2-c\frac{\sigma}{\sqrt{M}}=\alpha/2-c\frac{\sigma}{\sqrt{M}}=\alpha/2 - \eta\alpha,$$ which leads to Equation \eqref{volpertrade}. Importantly, this simple cost analysis and Equation \eqref{volpertrade} derived from it enable us to design simple prediction formulas for $\eta$ after a change in the tick value.

\subsection{Prediction of the cost of market and limit orders}
\label{sec:predictionmodel}

Based on the fact that Equation \eqref{volpertrade} should hold for a large tick asset for any tick value, the authors in \citet*{dayri2012large} establish three prediction formulas for the new value of $\eta$, and therefore for the new cost of market and limit orders, after a change in the tick value of an asset. Each of the three formulas corresponds to different assumptions. For simplicity, we only present here a formula which does not require any prior regression analysis and assumes a linear shape for the cumulative latent liquidity.\\

Let us consider a large tick asset for which the current tick value is $\alpha_0$ and associated is $\eta_0$. Then, if the tick value is changed to $\alpha$, 
%according to the prediction formula in \citet*{dayri2012large}, 
the formula for the new parameter $\eta$ gives:

\begin{equation}
\eta \sim (\eta_0 + 0.1)(\frac{\alpha_0}{\alpha})^{1/2} - 0.1. \label{v2}
\end{equation}

We now comment this formula and its use to predict the new microstructural features of an asset after a tick value change:
\begin{itemize}
\item Formula \eqref{v2} actually holds only provided the asset remains a large tick asset after the change in the tick value. However, due to the concurrence  
 mechanism, market makers maintain a spread equal to one tick as long as they make profit from it, that is as long as $\eta<1/2$. The value of $\eta$ in the formula being decreasing with $\alpha$, we get that Formula \eqref{v2} holds provided $\alpha\geq \alpha^*$ with
 $$\alpha^*=\big(\frac{\eta_0+0.1}{0.6}\big)^{2}\alpha_0.$$
\item Formula \eqref{v2} enables us to tell whether the asset remains a large tick asset after the tick value change: if the forecast value of $\eta$ is greater than $1/2$ (that is $\alpha<\alpha^*$), the asset is predicted to become a small tick asset after the tick value modification. However, note that in that case, the forecast value of $\eta$ cannot really be interpreted beyond this (becoming small tick or not).
\item If the predicted value of $\eta$ is smaller than $1/2$, Formula \eqref{v2} provides the estimated $\eta$ after the tick value change, and therefore the estimated cost of market and limit orders. In particular, this allows us to tell ex ante whether a stock will become/remain favorable for market makers or exhibit balanced trading costs. This is probably the most relevant viewpoint in term of regulation.  
\end{itemize}

\subsection{What is a suitable tick value?}

From a regulatory perspective, a tick value can probably be seen as suitable if:

\begin{itemize}
\item The bid-ask spread is close to one tick, ensuring the presence of liquidity in the order book.
\item Transaction costs are close to zero for market orders. In that case, the market is efficient and market makers do not take advantage of the tick value to the detriment of final investors acting mainly as liquidity takers.  
\end{itemize}

Thus, in our approach, an asset enjoys a relevant tick value if it is a large tick asset and its $\eta$ parameter is close to $1/2$. Indeed, recall that the cost of a market order for a large tick asset is $\alpha/2-\eta\alpha$.\\

Note that according to Formula \eqref{v2}, starting from a large tick asset, the optimal tick value can be obtained setting $\alpha=\alpha^*$. With this optimality notion in mind, we conduct in the next section an empirical analysis of the Japanese pilot program on tick values.

\section{Analysis of the Tokyo Stock Exchange pilot program on tick values}
%\section{Data and Empirical Results} 
\label{empstudy}

\subsection{Data description} \label{secdata}

We use data from the 55 Japanese stocks of the TOPIX 100 index involved in the pilot program in 2014. Our database, provided by Capital Fund Management, records the time and price of every transaction, as well as the best bid and ask prices right before the transactions, from June 3, 2013 to December 30, 2014. We remove market data corresponding to the first and last hour of trading, as these periods have usually specific features due to the opening/closing auction. Three different phases are distinguished in this study: 
\begin{itemize}
\item Phase 0 (before the pilot program): from June 3, 2013 to January 13, 2014. 
\item Phase 1 (from the first implementation of the tick value reduction program to the second one): from January 14, 2014 to July 21, 2014. 
\item Phase 2 (after the second implementation of the tick value reduction program): from July 22, 2014 to December 30, 2014. 
\end{itemize}
The details of the pilot program are given in Table \ref{ticktable}.\\
\begin{table}[h!]\footnotesize
\begin{center}
\begin{tabular}{| r | r | r | r  | } \hline
Quoted price below (\textyen) & Phase 0 tick value (\textyen) & Phase 1 tick value (\textyen) & Phase 2 tick value (\textyen)   \\ \hline
1,000 & 1 & 1 & 0.1   \\ \hline
3,000 & 1 & 1 & 0.5   \\ \hline
5,000 & 5 & 1 & 0.5   \\ \hline
10,000 & 10 & 1 & 1  \\ \hline
30,000 & 10 & 5 & 5   \\ \hline
50,000 & 50 & 5 & 5  \\ \hline
100,000 & 100 & 10 & 10  \\ \hline
300,000 & 100 & 50 & 50  \\ \hline
500,000 & 500 & 50 & 50  \\ \hline
1,000,000 & 1000 & 100 & 100  \\ \hline
3,000,000 & 1000 & 500 & 500  \\ \hline
5,000,000 & 5000 & 500 & 500  \\ \hline
10,000,000 & 10000 & 1000 & 1000  \\ \hline
30,000,000 & 10000 & 5000 & 5000  \\ \hline
50,000,000 & 50000 & 5000 & 5000  \\ \hline
Higher prices & 100000 & 10000 & 10000   \\ \hline
\end{tabular}
\caption{Tick value reduction table.} \label{ticktable}
\end{center}
\end{table}

\subsection{Classification of the stocks in Phase 0}

In this work, we use the average spread in number of ticks, now denoted by $S$ and estimated as the average value of the bid-ask spread right before a transaction, to classify the stocks into three groups: 
\begin{itemize}
\item Small tick stocks: $S > 1.6$.
\item Large tick stocks: $S \leq 1.5$.
\item Ambiguous case between large and small tick: $1.5<S\leq 1.6$.
\end{itemize}

For large tick stocks, we use the parameter $\eta$ to distinguish between balanced stocks, for which market orders are reasonably costly, and stocks where market makers (viewed again as an aggregated class) obtain significant profit from liquidity takers thanks to the tick value. We use the following criterion:

\begin{itemize}
\item Balanced stocks: $\eta \geq 0.4$.
\item Market makers favorable stocks: $\eta<0.4$.
\end{itemize}

Note that small tick stocks will be considered balanced stocks. However, they do not fulfill the criteria for stocks having a suitable tick value, their spread being somehow too large.\\ 

For each of the 55 stocks, for Phase 0, we give in Table \ref{statsp0} the average spread ($S_0$), the value of the $\eta$ parameter ($\eta_0$) and tell whether the stock is large tick or not (Yes or No for the variable $\LT_0$) and whether it is balanced or not (Yes or No for the variable $\BAL_0$).\\

\begin{table}\footnotesize
\begin{center}
\vspace{-15mm}
\begin{tabular}{| c | c | c | c | c | } \hline
\bf Company Name & $S_0$ & $\eta_0$ & $\LT_0$ & $\BAL_0$ \\ 
                                &           &                & {\small Large Tick} & {\small Balanced}\\\hline \hline
Aeon Co Ltd & 1.07 & 0.18 & Yes & No   \\ \hline
ANA Holdings Inc & 1.23 & 0.09 & Yes & No   \\ \hline 
Asahi Class Co Ltd& 1.03 & 0.15 & Yes & No   \\ \hline 
Asahi Kasei Corp & 1.04 & 0.21 & Yes & No   \\ \hline 
Astellas Pharma Inc  & 1.05 & 0.14 & Yes & No  \\ \hline 
Bank of Yokohama Ltd & 1.02 & 0.23 & Yes & No   \\ \hline 
Canon Inc & 1.04 & 0.06 & Yes & No   \\ \hline 
Chubu Electric Power Co Inc & 1.25 & 0.49 & Yes & Yes   \\ \hline 
Daiwa Securities Group Inc & 1.04 & 0.20 & Yes & No   \\ \hline
Dalichi Sankyo Co Ltd & 1.15 & 0.32 & Yes & No   \\ \hline
Dai-ichi Life Insurance Co Ltd  & 1.10 & 0.22 & Yes & No \\ \hline 
Fujitsu Ltd  & 1.03 & 0.13 & Yes & No  \\ \hline
Hitachi Ltd & 1.04 & 0.10 & Yes & No   \\ \hline 
Honda Motor Co Ltd  & 1.04 & 0.10 & Yes & No \\ \hline
Inpex Corp & 1.27 & 0.55 & Yes & Yes   \\ \hline 
ITOCHU Corp& 1.05 & 0.18 & Yes & No  \\ \hline 
Japan Tobacco Inc & 1.04 & 0.12 & Yes & No   \\ \hline
JX Holdings Inc & 1.01 & 0.08 & Yes & No   \\ \hline 
Kansai Electric Power Co Inc  & 1.24 & 0.33 & Yes & No  \\ \hline 
Kirin Holdings Co Ltd & 1.43 & 0.53 & Yes & Yes   \\ \hline 
Kubota Corp & 1.34 & 0.46 & Yes & Yes  \\ \hline 
Komatsu Ltd & 1.30 & 0.29 & Yes & No   \\ \hline 
Marubeni Corp  & 1.02 & 0.12 & Yes & No  \\ \hline 
Mitsubishi Chemical Holdings Corp & 1.01 & 0.16 & Yes & No  \\ \hline 
Mitsubishi Corp & 1.22 & 0.36 & Yes & No   \\ \hline 
Mitsubishi Electric Corp  & 1.10 & 0.31 & Yes & No  \\ \hline 
Mitsubishi Estate Co Ltd & 1.70 & 0.60 & No & Yes   \\ \hline
Mitsubishi Heavy Industries Ltd & 1.01 & 0.11 & Yes & No   \\ \hline
Mitsubishi UFJ Financial Group Inc& 1.03 & 0.04 & Yes & No   \\ \hline
Mitsui Co Ltd & 1.12 & 0.23 & Yes & No   \\ \hline 
Mitsui Fudosan Co Ltd & 1.07 & 0.23 & Yes & No   \\ \hline 
Mizuho Financial Group Inc& 1.14 & 0.07 & Yes & No   \\ \hline 
Nissan Motor Co Ltd & 1.06 & 0.14 & Yes & No   \\ \hline 
Nippon Steel Sumitomo Metal Corp & 1.02 & 0.05 & Yes & No   \\ \hline 
Nippon Telegraph Telephone Corp & 1.04 & 0.08 & Yes & No  \\ \hline 
Nomura Holdings Inc & 1.05 & 0.06 & Yes & No   \\ \hline 
NTT DoCoMo Inc & 1.28 & 0.24 & Yes & No   \\ \hline 
ORIX Corp & 1.21 & 0.33 & Yes & No   \\ \hline 
Osaka Gas Co Ltd & 1.04 & 0.16 & Yes & No  \\ \hline
Panasonic Corp  & 1.14 & 0.16 & Yes & No  \\ \hline 
Resona Holdings Inc  & 1.00 & 0.07 & Yes & No  \\ \hline 
Ricoh Co Ltd & 1.13 & 0.36 & Yes & No   \\ \hline 
Seven I Holdings Co Ltd  & 1.06 & 0.16 & Yes & No  \\ \hline
Softbank Corp & 1.05 & 0.06 & Yes & No   \\ \hline 
Sony Corp & 1.17 & 0.24 & Yes & No   \\ \hline 
Sumitomo Corp  & 1.04 & 0.18 & Yes & No  \\ \hline 
Sumitomo Electric Industries Ltd & 1.24 & 0.40 & Yes & Yes   \\ \hline 
Sumitomo Mitsui Trust Holdings Inc & 1.01 & 0.14 & Yes & No   \\ \hline 
Sumitomo Mitsui Financial Group Inc & 1.15 & 0.08 & Yes & No   \\ \hline 
Takeda Pharmaceutical Co Ltd & 1.06 & 0.13 & Yes & No   \\ \hline 
Tokyo Gas Co Ltd & 1.05 & 0.16 & Yes & No   \\ \hline 
Tokio Marine Holdings Inc & 1.05 & 0.18 & Yes & No  \\ \hline
Toray Industries Inc & 1.03 & 0.13 & Yes & No   \\ \hline 
Toshiba Corp  & 1.03 & 0.06 & Yes & No  \\ \hline 
Toyota Motor Corp & 1.03 & 0.04 & Yes & No   \\ \hline 
 \end{tabular}
\caption{Average spread, value of $\eta$ and categories (large tick or not; balanced or not) for the 55 stocks in Phase 0.} \label{statsp0}
\end{center}
\end{table}

We see that all the assets but one (Mitsubishi Estate Co Ltd) are large tick stocks in Phase 0. However, among the remaining 54 large tick stocks, only five of them are balanced: Chubu Electric Power Co Inc, Inpex Corp, Kirin Holdings Co Ltd, Kubota Corp and Sumitomo Electric Industries Ltd. According to our framework, no tick value modification was necessary for these five assets. However, a tick value reduction can be beneficial for the 49 other stocks, which somehow justifies the will of the TSE to launch the pilot program.

\subsection{Phase 0 - Phase 1}
We now test the prediction formula \eqref{v2} between Phase 0 and Phase 1. We first select 12 stocks among the 55 stocks based on the following criteria: 

\begin{itemize}
\item These stocks are large tick assets during Phase 0.  
\item These stocks are involved in the tick value reduction program during Phase 1.  
\item For every stock, days on which multiple tick values are used\footnote{We recall that the tick value of a stock depends on its price, see Table \ref{ticktable}.} are removed from the database. We then choose the tick value right before the end of Phase 0 and the tick value right after the beginning of Phase 1 as two reference tick values. Stocks for which the numbers of days when the tick value is equal to its reference value in Phase 0 and Phase 1 are both greater than 10 are finally selected.
\end{itemize}

Twelve assets are remaining after this selection. For each of these stocks, based on the value of $\eta$ in Phase 0 ($\eta_0$), we predict the new value of $\eta$ in Phase 1 ($\eta_1^p$) using Formula \eqref{v2}. We provide confidence intervals based on the 25\% and 75\% quantiles of the distribution of the estimated daily $\eta$ in Phase 0\footnote{The estimated $\eta$ being the average of the daily estimations, in very few cases, the prediction can fall out of the confidence interval. Then we replace the prediction by the closest bound in the confidence interval.}. We also forecast whether the asset will be large tick in Phase 1 ($\LT_1^p$) and balanced in Phase 1 ($\BAL_1^p$). More precisely, considering a predicted $\eta$ larger than $1/2$ corresponds to an increase of the spread (recall that the situation $\eta>1/2$ is not compatible with a one tick spread):
\begin{itemize}
\item If $\eta_1^p \geq 0.55$, the asset is predicted to become a small tick asset after the tick value change.  
\item If $\eta_1^p < 0.5$, the asset is predicted to remain a large tick asset after the tick value change, with the forecast value for the new $\eta$ being meaningful and given by $\eta_1^p$. 
\item We qualify the situation $0.5\leq \eta_1^p<0.55$ as an ``ambiguous'' case between large tick and small tick.  
\end{itemize}

We compare the predictions to the actual quantities in Phase 1: $\eta_1$, $\LT_1$ and $\BAL_1$. The results are given in Table \ref{statsp1}.\\

\begin{table}\footnotesize\hspace{-9mm}
%\begin{center}
\begin{tabular}{| c | c| c || c | c | c | c || c | c | c |}\hline
\bf Company name & $S_0$ & $\eta_0$ & $S_1$ & $\eta_1$ & $\LT_1$ & $\BAL_1$ & $\eta_1^p$ & $\LT_1^p$ & $\BAL_1^p$ \\ \hline \hline
**Astellas Pharma Inc                       & 1.05 & 0.14& 1.72 & 0.43 & No & Yes & 0.66 [0.50,0.71] & No & Yes \\ \hline
**Canon Inc                                       & 1.04 & 0.06& 1.13 & 0.23 & Yes & No & 0.26 [0.19,0.27] & Yes & No \\ \hline 
**Honda Motor Co Ltd                       & 1.04 & 0.10& 1.23 & 0.32 & Yes & No & 0.34 [0.26,0.37] & Yes & No \\ \hline 
**Japan Tobacco Inc                          & 1.04 & 0.12& 1.23 & 0.32 & Yes & No & 0.39 [0.26,0.41] & Yes & No \\ \hline 
**Mitsui Fudosan Co Ltd                    & 1.07 & 0.23& 1.95 & 0.66 & No & Yes & 0.63 [0.52,0.69] & No & Yes \\ \hline
*Nippon Telegraph Telephone Corp & 1.04 & 0.08 & 2.00 & 0.62 & No & Yes & 0.46 [0.35,0.51] & Yes & Yes \\ \hline 
(*)*Seven I Holdings Co Ltd                & 1.06 & 0.16& 1.55 & 0.51 & \ambig & Yes & 0.49 [0.38,0.55] & Yes & Yes \\ \hline 
*Softbank Corp                                & 1.05 & 0.06& 1.85 & 0.50 & No & Yes & 0.40 [0.32,0.40] & Yes & Yes \\ \hline 
*Sumitomo Mitsui Financial Group Inc & 1.15 & 0.08& 1.33 & 0.34 & Yes & No & 0.47 [0.27,0.47] & Yes & Yes \\ \hline 
**Takeda Pharmaceutical Co Ltd       & 1.06 & 0.13& 1.46 & 0.43 & Yes & Yes & 0.42 [0.28,0.45] & Yes & Yes \\ \hline 
(*)*Tokio Marine Holdings Inc               & 1.05 & 0.18& 1.39 & 0.46 & Yes & Yes & 0.53 [0.41,0.57] & \ambig & Yes \\ \hline 
**Toyota Motor Corp                        & 1.03 & 0.04 & 1.36 & 0.32 & Yes & No & 0.35 [0.27,0.33] & Yes & No \\ \hline 
\end{tabular}
\caption{For the 12 selected stocks: Average spread and value of $\eta$ in Phase 0 and Phase 1, categories in Phase 1, and predictions for $\eta$ and the categories. The number of stars * in front of a company name represents the number of good predictions (one for being large tick or not, one for being balanced or not). A star between brackets (*) corresponds to an ``ambiguous'' case.} \label{statsp1}
%\end{center}
\end{table}

The obtained average relative prediction error for $\eta_1$, that is the average of the $|\eta_1^p-\eta_1|/\eta_1$ is less than 18\%. This shows that thanks to Formula \eqref{v2}, we can forecast the new value of $\eta$, and therefore the new trading costs, with a good accuracy. Our prediction for $\eta$ being quite sharp, it is no surprise that we are able to forecast whether or not a stock is going to remain large tick and whether it is balanced in Phase 1. For nine of the stocks, our forecast was that it would remain large tick, and seven of these predictions were correct (classifying the ``ambiguous'' case as correct). We also predicted that two of the assets would become small tick and both predictions were correct. Thus, the predictive power of our methodology is very high.\\

According to our optimality notion, three of the twelve stocks now enjoy a suitable tick value (balanced large tick stocks): Seven I Holdings Co Ltd, Takeda Pharmaceutical Co Ltd and Tokio Marine Holdings Inc. Such results could have been obtained ex ante using our approach. Indeed, following our methodology, a regulator or an exchange can anticipate the suitable way to operate a tick value change, especially when the goal is to decrease the tick value of an unbalanced stock.\\

\begin{figure}
\centering
\includegraphics[width=\textwidth,height=8cm]{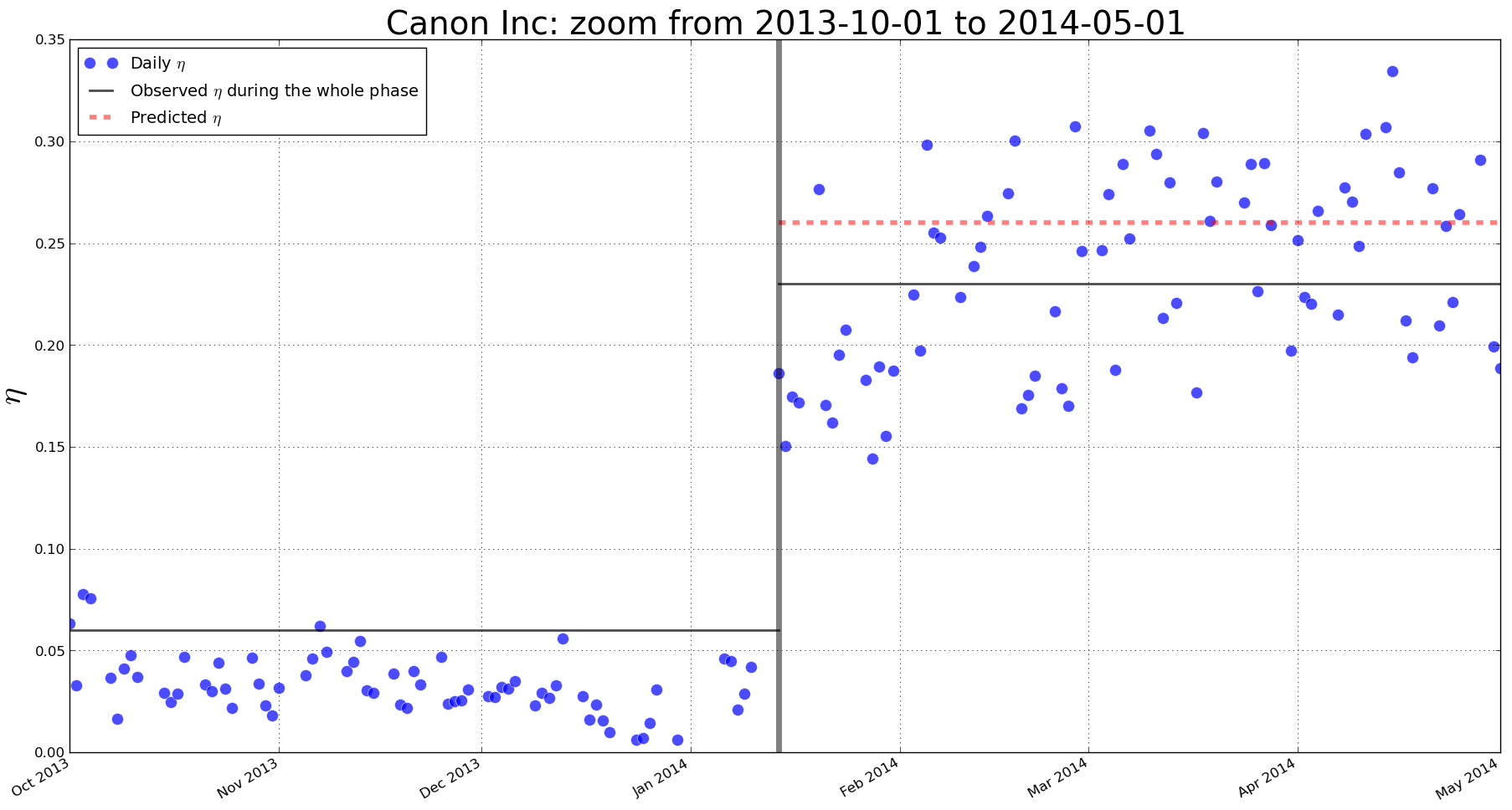}
\caption{Daily estimations of $\eta$ for the last 3.5 months of Phase 0 and the first 3.5 months of Phase 1 for the stock Canon Inc.}\label{fig:Canon_history}
\end{figure}

To end this subsection, as an illustration, we give in Figure \ref{fig:Canon_history} detailed results about $\eta$ for the stock Canon Inc. More precisely, we provide daily estimations of $\eta$ for the last 3.5 months of Phase 0 and the first 3.5 months of Phase 1. We also add the average values of $\eta$ during both phases together with our forecast for the value of $\eta$ in Phase 1. We see on this example that our prediction is very close to the realized value.

\subsection{Phase 1 - Phase 2}

The tick value reduction program affects much more stocks in Phase 2. Using the same selection criteria as previously (replacing Phase 0 by Phase 1 and Phase 1 by Phase 2), we find 48 stocks that are large tick assets during Phase 1 and have their tick value effectively reduced in Phase 2. We draw a similar analysis as the one for Phase 0-Phase 1. The results are given in Table \ref{statsp2}, where the index $1$ is used to denote quantities in Phase 1 and the index $2$ quantities in Phase 2.\\

\begin{table}\footnotesize\hspace{-9mm}
\begin{tabular}{| c | c| c || c | c | c | c || c | c | c |} \hline
\bf Company Name & $S_1$ & $\eta_1$ & $S_2$ & $\eta_2$ & $\LT_2$ & $\BAL_2$ & $\eta_2^p$ & $\LT_2^p$ & $\BAL_2^p$ \\ \hline \hline
**Aeon Co Ltd & 1.03 & 0.12 & 1.16 & 0.16 & Yes & No & 0.21 [0.18,0.26] & Yes & No \\ \hline 
**Asahi Class Co Ltd & 1.01 & 0.14 & 2.49 & 0.87 & No & Yes & 0.67 [0.52,0.82] & No & Yes \\ \hline
**Asahi Kasei Corp & 1.03 & 0.23 & 3.43 & 1.02 & No & Yes & 0.93 [0.76,1.07] & No & Yes \\ \hline 
**ANA Holdings Inc & 1.00 & 0.02 & 1.37 & 0.30 & Yes & No & 0.29 [0.26,0.31] & Yes & No \\ \hline 
**Bank of Yokohama Ltd & 1.02 & 0.22 & 3.28 & 0.98 & No & Yes & 0.92 [0.77,1.05] & No & Yes \\ \hline 
(*)Canon Inc & 1.13 & 0.23 & 1.59 & 0.43 & \ambig & Yes & 0.36 [0.30,0.43] & Yes & No \\ \hline 
**Chubu Electric Power Co Inc & 1.10 & 0.31 & 1.45 & 0.44 & Yes & Yes & 0.48 [0.40,0.55] & Yes & Yes \\ \hline 
**Daiwa Securities Group & 1.02 & 0.19 & 3.42 & 1.09 & No & Yes & 0.81 [0.70,0.90] & No & Yes \\ \hline 
**Dai-ichi Life Insurance Co Ltd & 1.08 & 0.25 & 1.35 & 0.35 & Yes & No & 0.39 [0.33,0.47] & Yes & No \\ \hline 
**Dalichi Sankyo Co Ltd & 1.10 & 0.27 & 1.43 & 0.40 & Yes & Yes & 0.43 [0.35,0.50] & Yes & Yes \\ \hline 
**Fujitsu Ltd& 1.01 & 0.16 & 2.80 & 0.97 & No & Yes & 0.72 [0.60,0.80] & No & Yes \\ \hline 
(*)*Hitachi Ltd & 1.01 & 0.09 & 2.55 & 0.81 & No & Yes & 0.50 [0.42,0.58] & \ambig & Yes \\ \hline 
*Honda Motor Co Ltd & 1.23 & 0.32 & 1.69 & 0.49 & No & Yes & 0.49 [0.43,0.55] & Yes & Yes \\ \hline
*Inpex Corp & 1.08 & 0.25  & 1.34 & 0.37 & Yes & No & 0.40 [0.35,0.45] & Yes & Yes \\ \hline 
**ITOCHU Corp& 1.03 & 0.13 & 1.17 & 0.24 & Yes & No & 0.23 [0.18,0.27] & Yes & No \\ \hline 
(*)*Japan Tobacco Inc& 1.23 & 0.32 & 1.86 & 0.55 & No & Yes & 0.50 [0.40,0.57] & \ambig & Yes \\ \hline
(*)*JX Holdings Inc  & 1.01 & 0.07 & 1.52 & 0.41 & \ambig & Yes & 0.44 [0.38,0.50] & Yes & Yes \\ \hline 
**Kansai Electric Power Co Ltd & 1.07 & 0.25 & 4.20 & 1.03 & No & Yes & 1.01 [0.86,1.17] & No & Yes \\ \hline 
*Kirin Holdings Co Ltd & 1.10 & 0.29 & 1.31 & 0.34 & Yes & No & 0.45 [0.31,0.57] & Yes & Yes \\ \hline 
*Komatsu Ltd & 1.10 & 0.24  & 1.50 & 0.46 & Yes & Yes & 0.39 [0.33,0.43] & Yes & No \\ \hline 
**Kubota Corp & 1.15 & 0.37 & 1.69 & 0.59 & No & Yes & 0.57 [0.49,0.64] & No & Yes \\ \hline 
**Marubeni Corp & 1.01 & 0.10 & 1.99 & 0.57 & No & Yes & 0.54 [0.44,0.62] & No & Yes \\ \hline 
*Mitsubishi Chemical Holdings  & 1.00 & 0.08 & 1.97 & 0.59 & No & Yes & 0.46 [0.36,0.52] & Yes & Yes \\ \hline
**Mitsubishi Corp & 1.05 & 0.18 & 1.40 & 0.36 & Yes & No & 0.29 [0.24,0.34] & Yes & No \\ \hline 
**Mitsubishi Electric Corp & 1.08 & 0.29 & 1.50 & 0.49 & Yes & Yes & 0.45 [0.39,0.52] & Yes & Yes \\ \hline 
**Mitsubishi Estate Co Ltd & 1.48 & 0.53 & 2.46 & 0.84 & No & Yes & 0.79 [0.71,0.89] & No & Yes \\  \hline 
**Mitsubishi Heavy Industries & 1.01 & 0.11 & 2.42 & 0.80 & No & Yes & 0.56 [0.45,0.66] & No & Yes \\ \hline 
**Mitsubishi UFJ Financial Group Inc & 1.00 & 0.03 & 1.44 & 0.32 & Yes & No & 0.30 [0.28,0.32] & Yes & No \\ \hline 
**Mitsui Co Ltd & 1.04 & 0.14 & 1.18 & 0.21 & Yes & No & 0.24 [0.20,0.28] & Yes & No \\ \hline 
**Nippon Steel Sumitomo Metal Corp& 1.00 & 0.03 & 1.29 & 0.31 & Yes & No & 0.30 [0.27,0.34] & Yes & No \\ \hline
(*)*Nissan Motor Co Ltd & 1.01 & 0.09 & 2.27 & 0.62 & No & Yes & 0.50 [0.42,0.58] & \ambig & Yes \\ \hline 
Nomura Holdings Inc& 1.00 & 0.05  & 1.90 & 0.51 & No & Yes & 0.36 [0.33,0.40] & Yes & No \\ \hline 
**NTT DoCoMo Inc & 1.03 & 0.17 & 1.28 & 0.34 & Yes & No & 0.28 [0.24,0.32] & Yes & No \\ \hline 
**ORIX Corp & 1.06 & 0.23 & 1.23 & 0.33 & Yes & No & 0.37 [0.32,0.42] & Yes & No \\ \hline 
**Osaka Gas Co Ltd & 1.00 & 0.12 & 2.21 & 0.81 & No & Yes  & 0.59 [0.47,0.70] & No & Yes \\ \hline
**Panasonic Corp & 1.03 & 0.14 & 1.19 & 0.22 & Yes & No & 0.24 [0.20,0.28] & Yes & No \\ \hline
*Resona Holdings Inc & 1.00 & 0.06 & 1.84 & 0.56 & No & Yes & 0.41 [0.36,0.45] & Yes & Yes \\ \hline
**Ricoh Co Ltd & 1.05 & 0.25 & 1.23 & 0.29 & Yes & No & 0.39 [0.33,0.46] & Yes & No \\ \hline 
**Sony Corp  & 1.04 & 0.16 & 1.49 & 0.37 & Yes & No & 0.26 [0.21,0.32] & Yes & No \\ \hline 
**Sumitomo Corp & 1.03 & 0.14 & 1.17 & 0.20 & Yes & No & 0.24 [0.19,0.29] & Yes & No \\ \hline 
*Sumitomo Electric Industries & 1.07 & 0.29 & 1.32 & 0.37 & Yes & No & 0.45 [0.36,0.53] & Yes & Yes \\ \hline 
(*)*Sumitomo Mitsui Financial Group Inc & 1.33 & 0.34 & 1.92 & 0.59 & No & Yes & 0.52 [0.46,0.59] & \ambig & Yes \\ \hline 
**Sumitomo Mitsui Trust Holdings Inc & 1.00 & 0.12 & 1.74 & 0.62 & No & Yes & 0.59 [0.48,0.67] & No & Yes \\ \hline 
**Takeda Pharmaceutical Co  & 1.46 & 0.43 & 2.19 & 0.67 & No & Yes & 0.65 [0.56,0.73] & No & Yes \\ \hline 
**Tokio Marine Holdings Inc & 1.39 & 0.46  & 2.10 & 0.70 & No & Yes & 0.70 [0.61,0.76] & No & Yes \\ \hline 
**Tokyo Gas Co Ltd & 1.01 & 0.14  & 2.54 & 0.91 & No & Yes & 0.66 [0.55,0.76] & No & Yes \\ \hline 
**Toray Industries Inc & 1.02 & 0.14 & 3.24 & 0.98 & No & Yes & 0.64 [0.51,0.77] & No & Yes \\ \hline
Toshiba Corp& 1.00 & 0.05 & 1.74 & 0.52 & No & Yes & 0.37 [0.31,0.42] & Yes & No \\ \hline 
 \end{tabular}
\caption{For the 48 selected stocks: Average spread and value of $\eta$ in Phase 1 and Phase 2, categories in Phase 2, and predictions for $\eta$ and the categories. The meaning of the stars in front of the company names is the same as for Table \ref{statsp1}.} \label{statsp2}
\end{table}
   
Once again, we obtain an excellent accuracy for predicting the value of $\eta$ in Phase 2 based on that in Phase 1. Indeed, the average relative error is here less than 17\%. Among the 48 assets, 16 of them are predicted to become small tick stocks and all these predictions are correct. Moreover, 28 stocks are predicted to remain large tick and 23 of these predictions are correct (taking the ambiguous cases as correct). Regarding the fact of being balanced, more than 85\% of our predictions are realised. Hence the study of the evolution of the market between Phase 1 and Phase 2 confirms what was found for Phase 0-Phase 1:  our device based on $\eta$ enables us to forecast ex ante the consequences of a change in the tick value on the market microstructure.\\ 

Note that after this second phase, the stocks Canon Inc, Chubu Electric Power Co Inc, Dalichi Sankyo Co Ltd, JX Holdings Inc, Komatsu Ltd and Mitsubishi Electric Corp seem to have a suitable tick value (balanced large tick stocks).

\section{Conclusion}\label{conclu}

Based on data from the TSE pilot program, we have studied the effects of tick value changes on the microstructure of large tick stocks. This has been done using the microstructural parameter $\eta$ which summarizes the high frequency features of a large tick asset, in particular the associated trading costs. The prediction formula suggested in \citet*{dayri2012large} for the new value of the parameter $\eta$ after a tick value change has been tested using all the stocks of the Japanese experiment. We have compared the prediction results in Phase 1 and Phase 2 with the realized $\eta$, and shown that Formula \eqref{v2} provides very accurate forecasts. In particular, we can predict ex ante whether a large tick stock will become a small tick stock after a tick value change and whether or not its associated trading costs will be balanced between market makers and liquidity takers.\\

This work validates the quantitative tools developed in \citet*{dayri2012large} for studying the consequences of a tick value modification. It provides detailed practical guidelines for market regulators and exchanges searching for optimal tick values. Indeed, it can help them choose suitable tick values without applying any trial and error method, which may largely reduce the duration and cost of pilot programs. 

\bibliography{ref_submit}

\end{document}